\documentclass[epj]{svjour}
\usepackage{amsmath}
\usepackage{amsmath}
\usepackage{amssymb}
\usepackage{amsfonts}
\usepackage{graphicx}
\usepackage{float}
\usepackage{wrapfig}
\usepackage{subfigure}
\usepackage{bm}
\usepackage{units}
\usepackage{grffile}
\usepackage{color}

\newcommand{\bk}{\boldsymbol{k}}
\newcommand{\bx}{\boldsymbol{x}}

\newcommand{\bu}{\boldsymbol{u}}
\newcommand{\bv}{\boldsymbol{v}}

\newcommand{\bhv}{\boldsymbol{\hat{v}}}

\newcommand{\bq}{\boldsymbol{q}}
\newcommand{\bp}{\boldsymbol{p}}

\def\ADDB#1{{\textcolor{black}{#1}}}     

\begin{document} \title{On the inverse energy transfer in rotating
turbulence} \author{Michele Buzzicotti\inst{1}, Patricio Clark Di
Leoni\inst{1} \and Luca Biferale\inst{1}}
\institute{Department of Physics \& INFN, University of Rome `Tor
Vergata', Via della Ricerca Scientifica 1, 00133 Rome, Italy}
\date{Received: date / Revised version: date} \abstract{ Rotating
turbulence is an example of a three-dimensional system in which an
inverse cascade of energy, from the small to the large scales, can be
formed. While usually understood as a byproduct of the typical
bidimensionalization of rotating flows, the role of the
three-dimensional modes is not completely comprehended yet. In order to
shed light on this issue, we performed direct numerical simulations of
rotating turbulence where the 2D modes falling in the plane
perpendicular to rotation are removed from the dynamical evolution. Our
results show that while the two-dimensional modes are key to the
formation of a stationary inverse cascade, the three-dimensional degrees
of freedom play a non-trivial role in bringing energy to the larger
scales also. Furthermore, we show that this backwards transfer of energy
is carried out by the homochiral channels of the three-dimensional
modes.  \\ {PACS-key}{discribing text of that key}   \and
{PACS-key}{discribing text of that key} } 
\maketitle

\let\thefootnote\relax\footnotetext{Postprint version of the article published on Eur. Phys. J. E (2018) 41: 131, DOI: 10.1140/epje/i2018-11742-4}

\section{Introduction}

In the classical picture of three dimensional turbulence, energy is
injected in the larger scales of the problem and then transferred to the
smaller ones in a process known as a direct energy cascade
\cite{Frisch,sagaut2018homogeneous,Pope}. Since Richardson's observations and
Kolmogorov's prediction turbulent cascades have been studied in many
systems, e.g. in rotating flows, stratified flows, and magnetohydrodynamics
flows \cite{DavidsonRot}.  One of the most important results so far has
been Kraichnan's prediction \cite{Kraichnan67} of the presence of an
{\it inverse} cascade of energy in two dimensional turbulence. Under
this regime, energy flows from small to large scales. Later, inverse
cascades have also been studied in three dimensional systems such as
rotating flows \cite{Smith96,Mininni09,Mininni09b}, shallow fluid layers
\cite{Nastrom84,Celani10}, oceanic flows \cite{Aluie17},
magnetohydrodynamics \cite{Alexakis06,Mininni07}, and helically
decimated flows \cite{Biferale12,Biferale13,Alexakis17}. While in
two dimensional turbulence the inverse energy transfer can be predicted
and understood by the presence of two positive definite quadratic inviscid
invariants, namely energy and enstrophy, the same argument cannot be
extended to the three dimensional systems, where the second quadratic
inviscid invariant, helicity, is not sign definite. Understanding how
inverse cascades are formed in three dimensional problems is the subject
of ongoing research.

In this work we focus on the case of rotating turbulence. Flows under
rotation present a rich phenomenology with plenty of physical interest,
moreover they are common in nature, e.g. in the atmosphere and in the
oceans \cite{Cho08}, \ADDB{in planetary cores \cite{Lereun17}}, as well
as in several engineering problems \cite{Dumitrescu04}. The presence of
the Coriolis force in these flows breaks isotropy \cite{Sen12},
generates inertial waves and gives rise to the formation of large scales
columnar vortices \cite{DavidsonRot,Biferale16}. The resulting flows
look almost bidimensional, with most of the energy accumulated in the
modes perpendicular to the rotation axis, as seen in simulations
\cite{Smith96,Horne13} and experiments \cite{Moisy11}.  These effects
happen because the nature of the nonlinear interactions is changed
\cite{Cambon89,Cambon97} with the appearance of resonant interactions
due to the action of the inertial waves in the turbulent flow
\cite{Nazarenko}. Resonant interactions are known to play an important
role in turbulent dynamics \cite{Newell69,Galtier03,Chen05}, and their
action has been studied directly in experiments
\cite{Yarom14,Campagne15} and simulations
\cite{Clark14a,Clark15c,Clark16a}. While it can be shown that the effect
of resonant interactions can make the energy transfer anisotropic, with
energy being preferentially transfered to modes closer and closer to the
plane perpendicular to the rotation axis, it can also be shown that they
cannot transfer energy directly into the perpendicular plane
\cite{Waleffe93}.  Furthermore, the modes perpendicular to the rotation
axis (the 2D modes, because of their two-dimensional nature) and the
resonant triads are decoupled under strong rotation \cite{Waleffe93}. So
the question of how the transfer of energy between the 2D and the rest
of the modes (the 3D modes, as they encompass all three-dimensional
modes in the flow) takes place remains open
\cite{Biferale16,Buzzicotti18}, with eddies \cite{Jacquin90,Moisy11} and
quasi-resonant interactions \cite{Alexakis15,Gallet15} appearing to have
an important role in the rotating turbulence dynamics.  \ADDB{Of
particular interest to the present work are the simulations done by
\cite{Lereun17}, where they showed that if the 2D modes are damped,
the system then enhances the creation of waves and small scale
structures, suggesting that the balance between 2D and 3D modes modes is
indeed delicate and dynamic. A similar experiment was also performed in
convective flows \cite{Calzavarini06}.} It is important to stress that
in numerical simulations on a finite-box domain exact resonant
interactions may be ``lost'' due to discretization effects at small
wavenumbers \cite{Bellet06}, thus making quasi-resonant interactions
even more important at the large scales \cite{Smith02,smith2005near}.

The phenomenon of bidimensionalization is an important one because the
inverse cascade in rotating flows is often seen as a byproduct of the
emerging 2D dynamics \cite{Smith96,Mininni09,Mininni09b}. In this
picture, the decoupled modes behave as purely 2D and Kraichnan's result
is recovered \cite{Kraichnan67}. As can be expected, wave
turbulence theories prohibit the formation of 2D solutions
\cite{Bellet06}. In these theories, the transfer of energy towards the
2D manifold becomes too weak to trigger an inverse cascade. It should be
noted, nonetheless, that it can be shown that bidimensionalization can
be achieved in periodic domains in the low Rossby limit
\cite{Gallet15}. All things considered, this scenario begs many
questions: if there is some coupling between the 2D and the 3D modes
that is able to put energy from the former to the latter, then wouldn't
this coupling also work the other way? If resonant interactions can make
the energy transfer anisotropic, aren't they also contributing to the
inverse transfer of energy? Moreover, it has been shown that even in
fully homogeneous and isotropic three dimensional turbulence, there are
channels that take energy backwards \cite{Biferale12,Biferale13}, that
the action of these channels can be enhanced in different geometries
\cite{Biferale17}, and that these channels can couple the 2D and 3D
modes in rotating turbulence \cite{Buzzicotti18}. This adds a further
avenue to explore.

In order to better understand the mechanisms behind the inverse cascade
in rotating turbulence we perform simulations where the two dimensional
modes are conservatively removed and compare them to a simulation of the
full Navier-Stokes equations. In this way the flow can never, by
construction, become two dimensional, therefore all the purely
three-dimensional effects involved in the inverse energy transfer come
into light. The results show that although a stationary inverse cascade
is not formed, energy is nonetheless transferred and accumulated in
modes larger than the ones where it is injected. A
pseudo-bidimensionalization takes place where energy is condensated in
the lowest wavenumbers close to the perpendicular plane, forming a quasi
2D flow. In these states, the homochiral channels of the energy flux
bring the energy to the large scales while the heterochiral ones bring
it to the small scales, with the two of them balancing out. In summary,
our results show that while the 2D modes are key to the formation of an
inverse cascade in rotating flows, the 3D modes play a non-negligible
role in the distribution of energy, making the overall dynamics very
rich.

The paper is organized as follows, in Sec.~\ref{sec:eqs} we introduce
the equations of rotating turbulence and several of its core concepts,
explain the process used to keep only the three-dimensional modes, and
give details on the simulations we perform and the different quantities
we analyze, in Sec.~\ref{sec:results} we present the results coming from
our numerical simulations, and in Sec.~\ref{sec:conc} we give concluding
remarks.

\section{Rotating turbulence equations}
\label{sec:eqs}

The governing equations for an incompressible fluid in a rotating frame
can be written as

\begin{equation}
    \begin{cases}
    \partial_t \bm{u} + \bm{\omega} \times \bm{u} +2{\bm \Omega} \times {\bm u}
    = - \nabla p + \nu\Delta{\bu}+ \bm{f} \\
    \nabla \cdot \bm{u} = 0,
    \end{cases}
    \label{eq:navierstokes_nohv}
\end{equation}
where $\nu$ is the kinematic viscosity, $\bm{f}$ is an external forcing,
the term $2{\bm \Omega} \times {\bm u}$ is the Coriolis force produced
by rotation, and $\bm \Omega = \Omega \hat{z}$ is the angular velocity
with frequency $\Omega$ around the rotation axis $\hat{z}$. The fluid
density is constant and absorbed into the definition of pressure $p$.

Taking the curl of the linearized form of
Eq.~\eqref{eq:navierstokes_nohv} in
the dissipation- and force-less regime (i.e., $\nu=0$ and $\bm{f}=0$)
yields

\begin{equation}
    \partial_t (\nabla \times \bm {u}) = 2 \left ( {\bm{ \Omega} \cdot
    \nabla} \right ) {\bm u}.
    \label{eq:linearizedRotatingNS}
\end{equation}
The general solution of this equation is given by a superposition of
waves of the form

\begin{equation}
    \bu(\bx,t) = \sum_{\bk, s_k} {\bf h}_{s_k}(\bk) e^{i[\bk \cdot \bx -
    \omega_{s_k}(\bk) t ]}
    \label{eq:helicalwavesolution}
\end{equation}
where $s_k=\pm$, ${\bf h}_{s_k}(\bm{k})$ are the orthogonal eigenmodes of the curl
operator, $i \bk \times {\bf h}_{s_k} = s_k k \,{\bf h}_{s_k}$ \cite{Greenspan68}, and the wave frequencies, $\omega_{s_k}$,
are given by the dispersion relation,

\begin{equation}
    \omega_{s_k}(\bk)=s_k 2 \Omega \frac{k_z}{|\bk|},
    \label{eq:dispersrelation}
\end{equation}
where $k_z$ is the direction of the rotation axis.  These are the
aforementioned inertial waves. It follows that, for each wavevector
there are two waves with opposite sign of helicity. The right-handed
wave propagating in the direction of $\bk$ and the left-handed wave
propagating in the $-\bk$ direction. Inertial waves also bring resonant
triads into play. It is well known that in
Eq.~\eqref{eq:navierstokes_nohv},
Fourier modes interact in triads satisfying $\bk+\bp+\bq=0$, where $\bk$,
$\bp$, and $\bq$ are the three wavevectors involved in the triad. The
presence of inertial waves adds a second condition

\begin{equation}
    \omega_{s_k}(\bk) + \omega_{s_p}(\bp) + \omega_{s_q}(\bq) = 0,
    \label{resonance}
\end{equation}
known as the resonance condition. Resonant interactions are very
important to the evolution of a rotating turbulent flow, but they do not
encompass all of the interactions that happen in it
\cite{Clark14a,Clark16a}.

The Reynolds and the Rossby numbers are the two non-dimensional parameters
which control the dynamic evolution of the flow. They can be written,
respectively, as

\begin{equation}
Re = \frac{U L_f}{\nu}, \qquad
Ro = \frac{U}{2\Omega L_f} ,
\end{equation}
where $L_f\sim 1/k_f$ is the forcing scale, and $U$ is the rms velocity at
the forcing scale.  The Rossby number represents the ratio between the
Coriolis force and inertial forces in the flow. In the limit of large
Rossby numbers, $Ro\gg1$, the flow can evolve freely under its own
internal dynamics without being influenced by rotation.  In the
$Ro\lesssim 1$ regime, we can expect to observe effects of rotation on
the flow.

\subsection{Rotating turbulence on a reduced Fourier set}

From the dispersion relation in Eq.~\eqref{eq:dispersrelation} it is
clear that all wavenumbers lying in the Fourier space plane,
$(k_x,k_y,k_z=0)$, perpendicular to the rotation axis, do not give rise to
inertial waves, as $\omega_{s_k}(\bk_{\perp}) = 0$. These are the
aforementioned 2D modes, with all the rest of the modes in the system
being the 3D modes. The two sets can be written explicitly in the
following way

\begin{align}
    \bk_{2D} &= \{\forall \bk \mid k_z = 0\},
    \\
    \bk_{3D} &= \{\forall \bk \mid k_z \neq 0\},
\end{align}
with $\bk_{2D}$ being the set of the 2D modes, and $\bk_{3D}$ the one of
the 3D modes. The 2D and 3D modes sets are sometimes referred to as the
``slow'' and ``fast'' manifolds, respectively
\cite{Greenspan69,Embid98,Smith99}.

As mentioned above, in turbulence under rotation energy tends to
accumulate in the 2D modes \cite{Mininni09b}. But it is not clear
whether the inverse cascade is produced only due to the action of these
modes, or if the 3D modes also play a role.  In this work, we directly
investigate the role of the 3D modes in the energy transfer. To do this,
we reduce the set of possible interactions described in the system of
Eqs.~\eqref{eq:navierstokes_nohv} to only the interactions which couple modes
inside the 3D set. Restricting the dynamics of
Eq.~\eqref{eq:navierstokes_nohv} to only the 3D modes can be accomplished by
using a generalized Galerkin projector, ${\cal P}$, which acts on the
velocity field as follows:

\begin{equation}
    {\bv}(\bx,t)= {\cal P} \, {\bu}(\bx,t)= \sum_{{\bk}} e^{i {\bk \cdot
    \bx}}\,\gamma_{\bk}\hat{\bu}(\bk,t)\,,
    \label{eq:decimOper}
\end{equation}
where ${\bv}(\bx,t)$ is the representation of the decimated velocity
field in the real space.  The factors $\gamma_{\bk}$ are chosen to be
either 1 or 0 with the following rule:
\begin{equation}
    \gamma_{\bk} =
    \begin{cases} 
        1, & \text{if } \bk \in \bk_{3D} \\
        0, & \text{if } \bk \in \bk_{2D} \,.
    \end{cases}
    \label{eq:theta}
\end{equation}
In this way the only active modes are the one inside the 3D set.
Moreover, the factors $\gamma_{\bk}$ preserve Hermitian symmetry
$\gamma_{\bk}= \gamma_{-\bk}$ so that ${\cal P}$ is a self-adjoint
operator. The resulting equations for the Fourier decimated velocity
field are then,

\begin{equation}
    \begin{cases}
    \partial_t {\bv} = {\cal P}[- {\bf \nabla}p -
    (\bv \cdot {\bf \nabla}  \bv )]  - 2{\bm \Omega} \times {\bv} + 
      \nu\Delta{\bv} +  {\cal P}\bm{f}\, \\ 
    \nabla \cdot \bv = 0.
    \end{cases}
    \label{eq:decimNS_nohv}
\end{equation}
In the above definition of the decimated equations, the nonlinear term
must be projected on the quenched decimated set, to constrain the
dynamical evolution to evolve on the same set of Fourier modes at all
times.  Moreover it is important to see that the resulting dynamics
still conserves total energy and helicity. Similarly, any initial
condition and the external forcing used must have a support on the same
decimated set of Fourier modes only.

\subsection{Energy spectra and fluxes}

We now define the different energy spectra and fluxes we use in this
study. The isotropic energy spectrum of the 3D modes can be written as

\begin{equation}
    E(k) = \frac{1}{2} \sum_{\substack{k \le|\bk| < k+1}}
    |\bm{\hat{v}}(\bk)|^2 .
    \label{eq:spettro}
\end{equation}
We can further decompose the spectrum into two components, one parallel
and one perpendicular

\begin{equation}
    e(k_\perp, k_\parallel) = \frac{1}{2} \sum_{k_\perp \leq |\bk \times
    \hat{z}| <  k_\perp +1 \atop k_\parallel \leq k_z < k_\parallel +
    1} |\bm{\hat{v}}(\bk)|^2,
    \label{eq:ekk}
\end{equation}
with $k_\perp = \sqrt{k^2_x + k^2_y}$ and $k_\parallel = k_z$. This
spectrum takes into account the anisotropic nature of the flow. When
plotting $e(k_\perp, k_\parallel)$, a trigonometric factor of $1/\sin\theta$, with
$\theta=\arctan{k_\parallel/k_\perp}$, will always be included.
Otherwise, even in the isotropic case (i.e., $\Omega=0$) the spectrum
\eqref{eq:ekk}
would not look isotropic, as there are many more modes with low $\theta$
entering in the summation.

The total energy flux has the form

\begin{equation}
    \Pi(k)=-\sum_{|\bk|\le k}  ik_j \hat{v}_i(-\bk)\sum_{\bp,\bq}
    \hat{v}_i(\bp)\hat{v}_j(\bq) \delta(\bp+\bq -\bk) .
    \label{eq:flusso}
\end{equation}
Notice that the decimated velocity $\bm{\hat{v}}(\bk)$ is non-zero only
when $\bk \in \bk_{3D}$, so the flux only takes into account the
interactions of the 3D modes. 

The energy flux can be further analyzed in terms of its homochiral and
heterochiral components. Doing this is important because it is known
that homochiral triads are responsible of opening a channel that takes
energy from the small to the large scales, even in 3D homogeneous and
isotropic turbulence \cite{Alexakis17,Sahoo18}. So in the context of
rotating turbulence it is interesting to see if rotation is only
producing a decoupling between 2D and 3D modes (whereby the inverse
cascade is then a product of the 2D dynamics), or if it also enhances
the backward flux produced by the homochiral interactions inside the 3D
manifold. To calculate the homo and heterochiral energy fluxes, we first
decompose the velocity field into the helical modes, ${\bf h_+}$
and ${\bf h_-}$, defined above, as proposed by
\cite{Waleffe92,Constantin88}, in order to obtain

\begin{align}
	\nonumber
    \bhv (\bk,t) &= \bhv^+(\bk,t) + \bhv^-(\bk,t)
    \\
    &= \hat v^+(\bk,t) {\bf h_+}(\bk) + \hat v^-(\bk,t) {\bf h_-}(\bk) \
    .
\end{align}
It is important to note that this decomposition can be performed for any
three dimensional incompressible field, not just for the case of
rotating flows. Under this change of basis, we can write the energy flux
corresponding to triads in which all modes have the same sign of the
helicity, the homochiral, and those in which one mode has a different
sign, the heterochiral. The corresponding homochiral ($\Pi^{\rm HO}(k)$)
and heterochiral ($\Pi^{\rm HE}(k)$) energy fluxes can then be defined as:

\begin{align}
    \nonumber
    \Pi^{\rm HO}(k)  =  - \sum_{|{\bk}| \le k, \atop
    \bq=\bk-\bp}
     [& \hat \bv^{+} (-\bk) \cdot
     (i\bk \cdot \hat \bv^+ (\bp)) \hat\bv^+ (\bq) + \\
     +& \hat \bv^{-} (-\bk) \cdot
     (i\bk \cdot \hat \bv^- (\bp)) \hat\bv^- (\bq)] \ ,
    \label{eq:flux_homo}
\end{align}

\begin{equation}
    \Pi^{\rm HE}(k) = \Pi(k)-\Pi^{\rm HO}(k) \ ,
    \label{eq:flux_hetero}
\end{equation}
where $\Pi(k)$ is the total energy flux defined in
Eq.~\eqref{eq:flusso}.

\subsection{Numerical simulations}
\label{sec:dataset}

As we are interested in the physics of the inverse cascade, we study
the flows defined above using hyperviscous dissipation, so as to reduce
the range of scales affected by viscosity. In this way,
Eqs.~\eqref{eq:navierstokes_nohv} become

\begin{equation}
    \begin{cases}
    \partial_t \bm{u} = - \nabla p - \bm{\omega} \times \bm{u} - 2{\bm \Omega} \times {\bm u} + \nu(-1)^{\alpha+1}\Delta^{\alpha}{\bu}+ \bm{f} \\
    \nabla \cdot \bm{u} = 0,
    \end{cases}
    \label{eq:navierstokes}
\end{equation}
and Eqs.~\eqref{eq:decimNS_nohv}

\begin{equation}
    \begin{cases}
    \partial_t {\bv} = {\cal P}[- {\bf \nabla}p -
    (\bv \cdot {\bf \nabla}  \bv )]  - 2{\bm \Omega} \times {\bv} + 
      \nu(-1)^{\alpha+1}\Delta^{\alpha}{\bv} +  {\cal P}\bm{f}\, \\ 
    \nabla \cdot \bv = 0.
    \end{cases}
    \label{eq:decimNS}
\end{equation}
\\
We perform direct numerical simulations of Eqs.~\eqref{eq:decimNS} in a
triple periodic domain with a fully dealiased parallel 3D pseudospectral
code using grids of up to $N^3 = 512^3$ collocation points. The time
integration has been performed with the second-order Adams-Bashforth
scheme with the viscous term integrated implicitly. The external
forcing, $\bm{f}$, is a delta correlated random process in
Fourier space
\begin{equation}
    \langle \hat{\bm{f}}({\bm{k}}) \hat{\bm{f}}^*({\bm{q}}) \rangle =
    F(\bm{k}) \delta_{\bm{k},\bm{q}} \hat{\bm{Q}}(\bm{k}),
\end{equation}
where $F(\bm{k})$ is an amplitude term that only has support around
$k_f$ and $\hat{\bm{Q}}(\bm{k})$ is a projector applied to guarantee
incompressibility.
The values of the different parameters used are
presented in Table~\ref{tbl:simulations}.  It is known that
hyper-viscosity introduces a bottleneck in the energy spectrum close to
the dissipative scales,  however, for the interests of this work, namely
the properties of the backward energy transfer from the forcing to the
large scales, we can safely assume that the spurious effects of
hyper-viscous dissipation are negligible. 

The simulations can be distinguished in two different sets. In the
first, PRJ-A, we fix the forcing properties and we study the system at
changing the rotation rate $\Omega$. In the second set, PRJ-B, we study
the effects of changing the forcing scale, keeping the same rotation
rate. In particular, in set PRJ-A, we keep the energy input fixed at
$k_f=30$, which guarantees a large enough inverse inertial range, and
vary $\Omega$ between $0$ and $160$. In set PRJ-B, instead, we fix
$\Omega=80$, which ensures a rotation rate strong enough to produce
backward cascade, and we change the input scales from $k_f=4$ up to
$k_f=50$. For this last simulation, we increase the resolution up to
$N^3=512^3$, in order to be able to force at $k_f=50$.  As a control and
benchmark, we also perform a simulation of the original non-decimated
system (Eqs.~\eqref{eq:navierstokes}), which we refer to as FULL. In
order to make a fair comparison with the other simulations, we only
force the 3D modes so as to not inject energy directly into the 2D ones.
More details about all the simulations are reported in
Table~\ref{tbl:simulations}. 

\begin{table}
    \begin{center}
    \begin{tabular}{|c|cccccc|}
    Simulation & Projected & $N$ & $\Omega$ & $k_{f}$ & $\varepsilon$ & $Ro$\\
    \hline
    FULL & No & $256$  & 80 & 30 & $0.045$ & $0.008$\\
    \hline
    PRJ-A1 & Yes & $256$  & 0 & 30 &  $0.06$ & $\infty$\\
    PRJ-A2 & Yes & $256$  & 5 & 30 & $0.06$ & $0.1$\\
    PRJ-A3 & Yes & $256$  & 15 & 30 & $0.06$ & $0.04$\\
    PRJ-A4 & Yes & $256$  & 40 & 30 & $0.06$ & $0.025$\\
    PRJ-A5 & Yes & $256$  & 80 & 30 & $0.06$ & $0.015$\\
    PRJ-A6 & Yes & $256$  & 160 & 30 & $0.06$ & $0.009$\\
    \hline
    PRJ-B1 & Yes & $256$  & 80 & 4 & $0.4$ & $0.007$ \\
    PRJ-B2 & Yes & $256$  & 80 & 15 & $0.1$ & $0.011$ \\
    PRJ-B3 & Yes & $256$  & 80 & 30 & $0.06$ & $0.015$\\
    PRJ-B4 & Yes & $512$  & 80 & 50 & $1.2$ & $0.03$\\
    \end{tabular}
    \end{center}
    \caption{Parameters used in the different simulations. ``Projected''
    indicates if the full equations (Eqs.~\eqref{eq:navierstokes}) or
    the projected ones (Eqs.~\eqref{eq:decimNS}) are used; $N$: number
    of collocation points in each spatial direction; $\Omega$: rotation
    rate; $k_{f}$: forced wavenumbers; $\varepsilon$: viscous energy
    dissipation; $Ro = (\varepsilon_f k_f^2)^{1/3}/\Omega$: Rossby
    number defined in terms of the energy injection properties. In all
    simulations the order of the Laplacian
    $(-1)^{\alpha+1}\nu\Delta^{\alpha}{\bv}$ is set to $\alpha=4$, and
    the kinematic viscosity to $\nu=1.8 \times 10^{-13}$, except for the
    simulation with resolution $N=512$, where the viscosity is set equal
    to $\nu=7.1 \times 10^{-14}$.}
    \label{tbl:simulations}
\end{table}

\section{Results}
\label{sec:results}

\begin{figure*}[h]
    \centering
    \includegraphics[width=.9\textwidth]{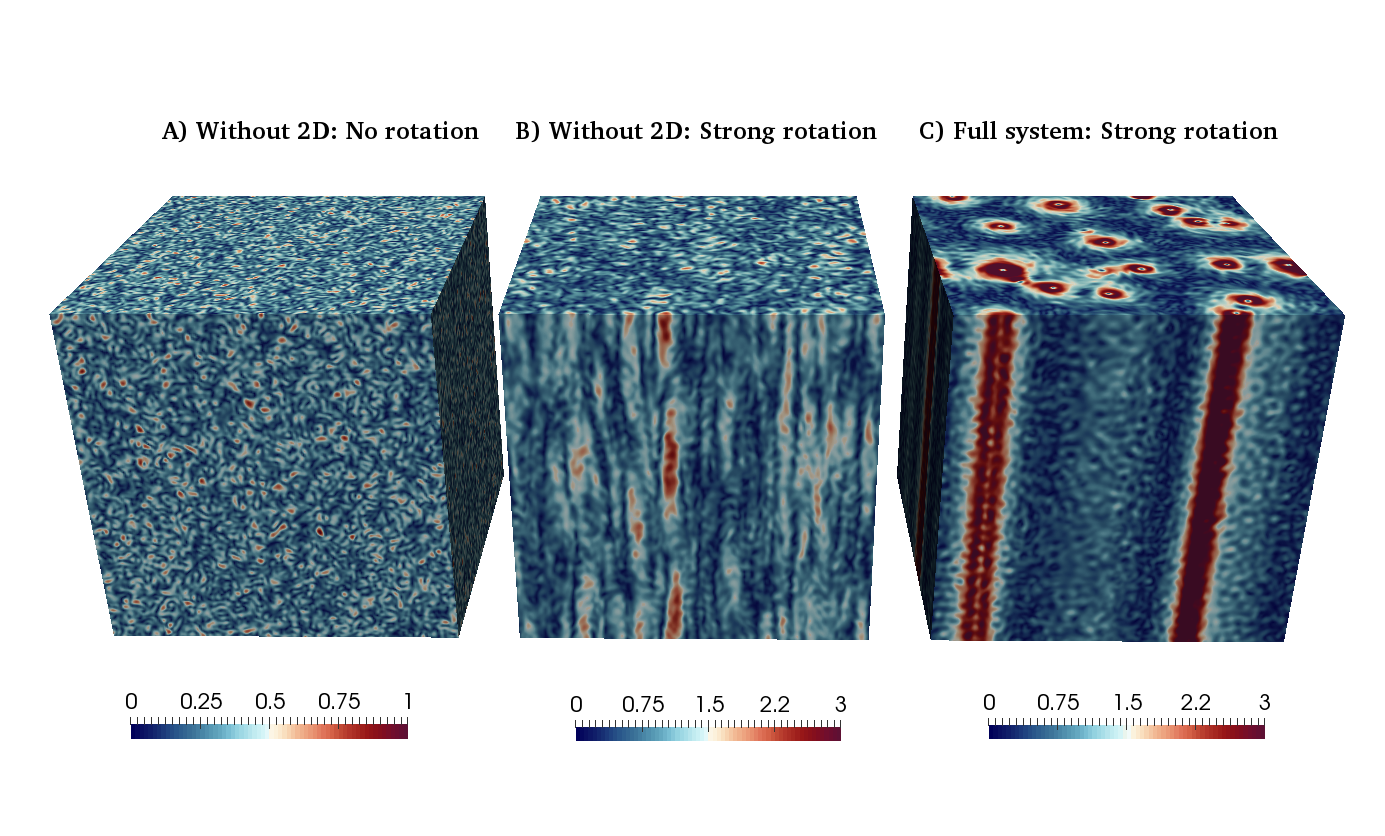}
    \caption{Visualizations of the absolute value of the velocity field
    for three simulations: A) projected system without the 2D manifold with no rotation (simulation PRJ-A1), B) projected system without the 2D manifold with strong rotation (simulation PRJ-A5), and C) the full system with strong rotation (simulation FULL).}
    \label{fig:viz}
\end{figure*}

In Fig.~\ref{fig:viz} we show visualizations of the absolute value of
the velocity for three simulations: PRJ-A1, PRJ-A5, and FULL (which have
$\Omega=0$, $\Omega=80$ and $\Omega=80$, and are shown in panels A, B
and C, respectively). As expected, simulation PRJ-A1 shows the typical
disordered structures found in homogeneous isotropic turbulence and
simulation FULL shows the characteristic columnar vortices with vertical
symmetry of rotating flows. On the other hand, PRJ-A5 shows vertical
structures that resemble the columnar vortices, but with no vertical
symmetry and with a stronger presence of disordered three-dimensional
structures. In a way, it is as if the system is trying to build the
columnar vortices but it is not able to successfully do it.

\begin{figure}[h]
    \centering
    \includegraphics[width=8.5cm]{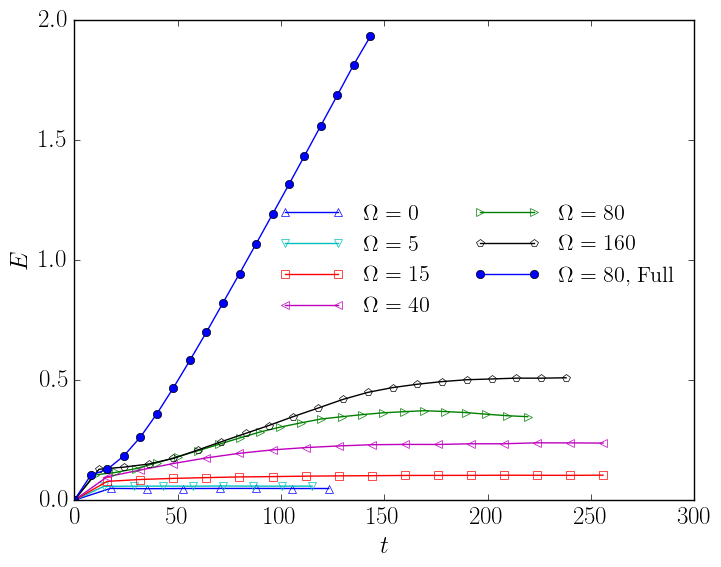}
    \caption{Evolution of total energy as a function of time for the set
    of simulations without the 2D manifold, PRJ-A, for different values of $\Omega$.
    For comparison the total energy evolution for the non-decimated
    (simulation FULL) system is also present (circular
    full markers).}
    \label{fig:energy}
\end{figure}

Moving on to a more quantitative analysis, in Fig.~\ref{fig:energy} we
present the total energy evolution for the set of decimated simulations
PRJ-A and for the simulation FULL, where we retain the 2D modes.  It is
evident that only keeping the 3D modes strongly affects the dynamical
evolution of the resulting systems. In particular, comparing the
evolution of FULL with that of PRJ-A5, we can see that in the case of
full Navier-Stokes equations the total energy increases with a constant
speed as a function of time, while in the decimated system the total
energy grows linearly only in a first transient of time, then it
saturates to a stationary state. This result suggests that without the
2D modes the system is not able to establish a backward energy transfer
stationary in time.  From the same Fig.~\ref{fig:energy} we can also
assess the effect of changing the Rossby number on the evolution of the
decimated systems. In particular we can see that if Rossby is large
enough, namely when $\Omega \leq 5$, the system does not seem to show a
transient period with constant energy increase, while this does happen
when $\Omega > 5$.  

\begin{figure}[h]
    \centering
    \includegraphics[width=8.5cm]{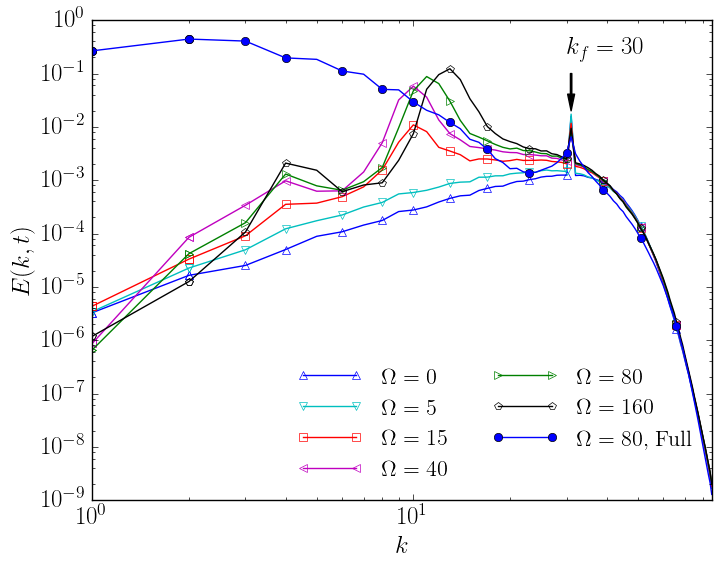}
    \caption{Energy spectra for the simulations without the 2D manifold, PRJ-A, at changing $\Omega$.
    For comparison the energy spectrum for the non-decimated
    (simulation FULL) system is also present (circular full markers).}
    \label{fig:spectra}
\end{figure}

In order to have a better understanding of how energy is distributed in
the system with 3D modes only, we show in Fig.~\ref{fig:spectra} the
energy spectra for the simulations in PRJ-A and for the simulation FULL.
The spectra of the decimated simulations is averaged on time once 
they reach their stationary regime, while the spectra from simulation
FULL is not averaged in time, as it never reaches a stationary state. If
rotation is not strong enough, energy is not transferred to the smaller
wavenumbers, as suggested in Fig.~\ref{fig:energy}. But if
rotation is strong, energy is indeed transferred to modes with $k<k_f$
even though there are no 2D modes in the system.  Interestingly, two
distinct peaks are formed around $10 \leq|\bk|\leq 12$ and around $5
\leq|\bk|\leq 7$. While the position of these peaks does not seem to be
greatly affected by the rotation rate, their amplitude is, with larger
values of $\Omega$ generating bigger peaks.  These peaked spectra differ
greatly from the equipartition spectrum $k^2$.  It is important to note
that there is no kind of large scale friction being used in these
simulations, so, as we will see below, the total energy flux of the
decimated cases in the region $k<k_f$ must be zero.

\begin{figure}[h]
    \centering
    \includegraphics[width=8.5cm]{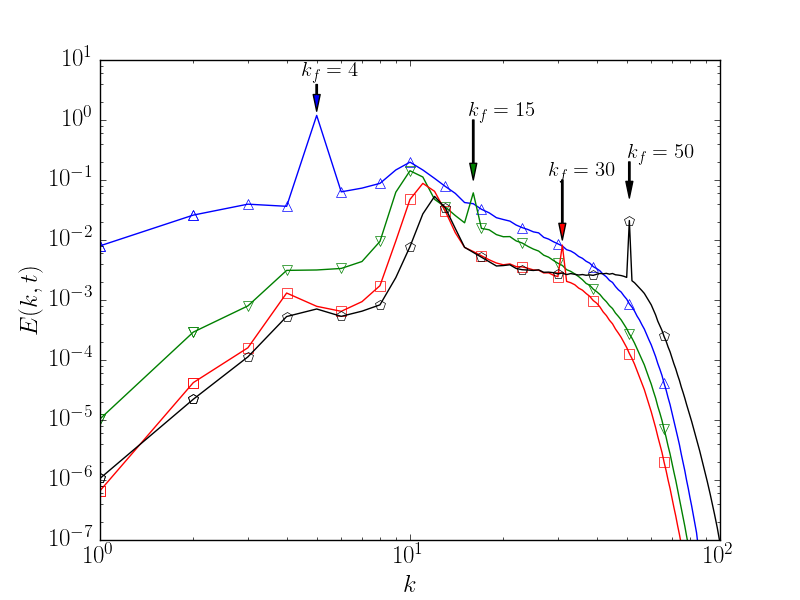}
    \caption{Energy spectra $E(k)$ of the simulations in set PRJ-B.}
    \label{fig:spectra_kf}
\end{figure}

Figure~\ref{fig:spectra_kf} shows the energy spectra of the simulations
in set PRJ-B, where we keep the rotation rate fixed at $\Omega=80$ and
vary the forcing scale $k_f$. In all cases, energy is accumulated around
the same peaks seen in Fig.~\ref{fig:spectra}, even for the case of
PRJ-B1, where the forcing is acting on wavenumbers smaller than those
were the peak is formed.

\begin{figure}[h]
    \centering
    \includegraphics[width=0.4\textwidth]{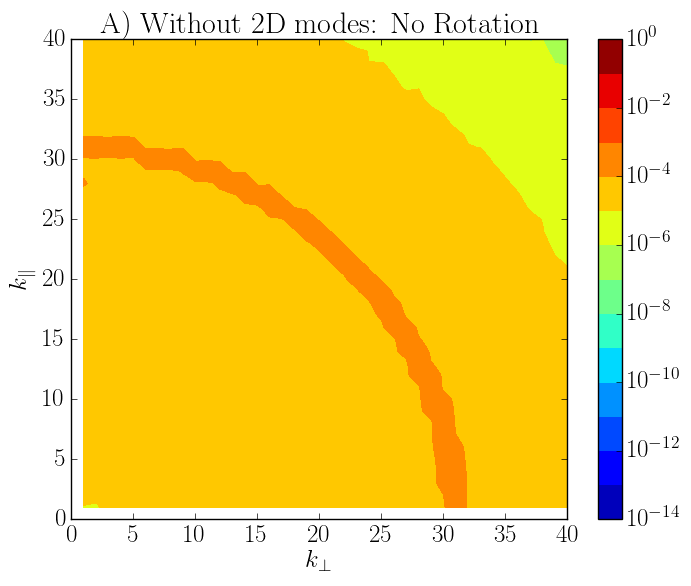}
    \includegraphics[width=0.4\textwidth]{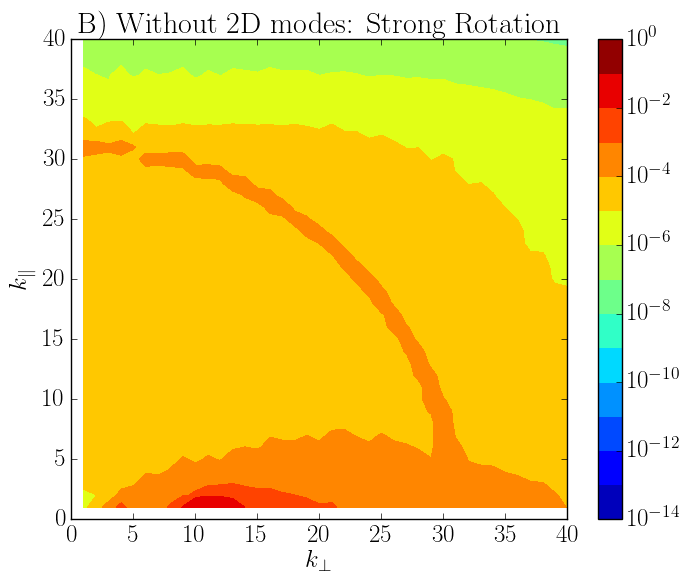}
    \includegraphics[width=0.4\textwidth]{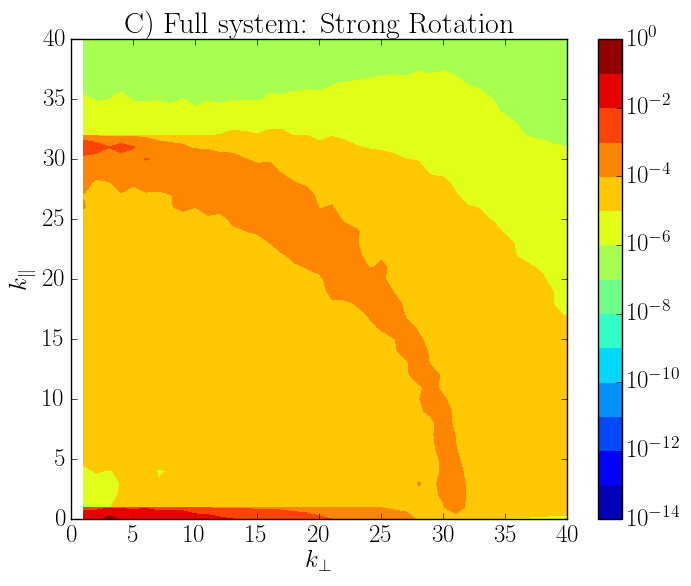}
    \caption{Decomposed energy spectra $e(k_\perp,k_\parallel)$ of
    simulations: A) PRJ-A1, B) PRJ-A5 and C) FULL.}
    \label{fig:ekk}
\end{figure}

So as to understand how the energy is distributed among modes parallel
and perpendicular to the rotation axis, in Fig.~\ref{fig:ekk}, we
analyze the decomposed energy spectra $e(k_\perp,k_\parallel)$ of the
simulations PRJ-A1, PRJ-A5, and FULL.  In the non-rotating case, PRJ-A1
shown in panel A, the spectral energy density forms concentric circles centered around $k=0$
in an isotropic fashion. On the other hand, when rotation is active this
distribution becomes anisotropic, with a stronger accumulation of energy
in modes with low $k_\parallel$, as expected from \cite{Waleffe93}.
While in simulation FULL (shown in panel C) energy is indeed located in
the 2D modes as commonly happens in rotating turbulence, in simulation
PRJ-A5 (shown in panel B) energy goes towards modes with low
$k_\parallel$ but is then squashed between $k=1$ and $k=5$, as it cannot
go to the 2D modes. So, the action of the 3D modes does take energy to
the larger scales and with a preference towards modes close to the 2D
modes.  It is important to note that in simulation FULL, no energy is
being injected directly into the 2D modes.  So while resonant
interactions make the spectra evolution anisotropic, quasi-resonant
interactions must be coming into play in order to couple the 3D and 2D
modes \cite{Smith02,Alexakis15,Gallet15,Clark16a}.

\begin{figure}[h]
    \centering
    \includegraphics[width=8.5cm]{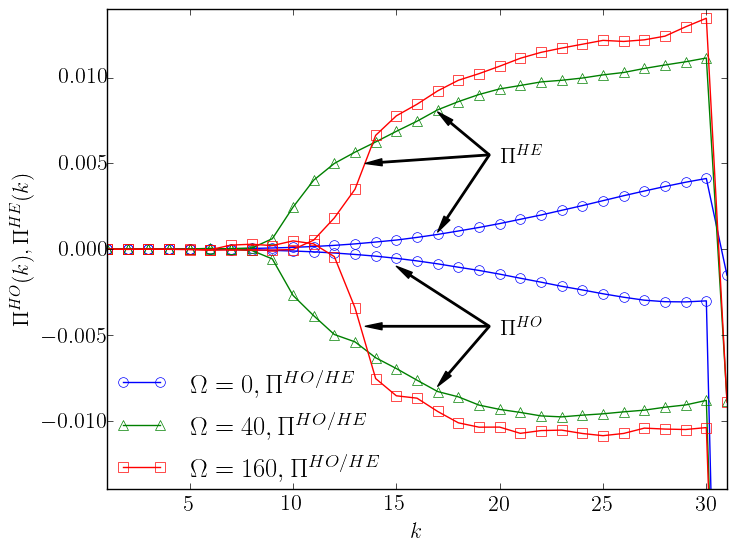}
    \caption{Homo and heterochiral contributions to the total energy
    flux, $\Pi^{HO}(k)$ and $\Pi^{HE}(k)$ respectively, of simulations
    PRJ-A1, PRJ-A3, and PRJ-A6.}
    \label{fig:fluxes}
\end{figure}

Finally, in order to understand how the stationary regimes are
sustained, we study the helical decomposition of the energy flux.  In
Fig.~\ref{fig:fluxes} we present the homochiral and heterochiral
contributions on the total energy flux of PRJ-A1, PRJ-A3 and PRJ-A6
(which have $\Omega=0$, $\Omega=40$ and $\Omega=160$, respectively). It
is interesting to observe that for the case of strong rotation the
stationary state is the result of the non-trivial cancellation of the
homo and heterochiral channels. So while the total flux is zero, this is
achieved by the dynamical balance of the channels that bring energy
forwards (the heterochiral) and the ones that bring it backwards
(the homochiral).  The same phenomenon has been observed in flows
composed by a combination of 2D3C (two dimension, three component) flows
\cite{Biferale17}. The amplitude and range of modes that are involved in
this flux balance does depend slightly on the rotation rate, with
simulation PRJ-A3 being the one with the largest number of modes that
had non-zero flux in the homo and heterochiral channels.

\section{Conclusions}
\label{sec:conc}

By performing simulations of rotating turbulence in which modes
perpendicular to the rotation axis (the 2D modes) were conservatively
removed we were able to assess the role of the 3D modes in the formation
of the inverse cascade. We showed that while a stationary inverse
cascade is not formed, energy is nonetheless transferred to the low
wavenumber modes.  The resulting non-trivial energy distribution is (i)
strongly anisotropic, concentrated close to the (removed) 2D plane and
(ii) highly peaked around $|\bk| \sim 10$.  Moreover, and more
importantly, we show that the stationary state is reached due to a
balance between homochiral and heterochiral transfers. The former
transferring energy backward and the latter forward.

In short, while the 2D modes are essential in order to have an inverse
cascade, the 3D modes play a non-negligible role distributing the energy
towards the large scales and in an anisotropic fashion.

The Fourier space decomposition in 2D-manifold and waves-component has
been performed also in small Rossby number turbulence confined in two
infinite walls perpendicular to the rotation axis, in this configuration
results show that the two-dimensional component has no effect on the
wave-component energetics \cite{scott2014wave}. Another study on
rotating turbulence in a triple-periodic domain, instead, claimed that
the backward energy cascade cannot be simplified as a 2D dynamics, but
it supports the picture that 3D-waves near resonant interactions
efficiently transfer energy from 3D modes to larger-scale 2D modes
\cite{smith2005near}.

It is important to comment about  a recent  study where a similar
problem is addressed   \cite{Lereun17} by damping the 2D slow modes
instead of decimating them    as we do here. By doing this, the presence
of  three dimensional inertial waves is enhanced and energy is
accumulated at small   scales, indicating a more efficient forward
energy cascade  once the 2D modes become unavailable  because of the
strong damping. Different from our case, in \cite{Lereun17} the
system is forced at large scale thus not allowing for  the energy to
flow backwards. Both works suggest that there is a non-trivial
correlation between 2D and 3D dynamics.

\section*{Acknowledgments} The research leading to these results has
received funding from the European Union's Seventh Framework Programme
(FP7/2007-2013) under grant agreement No.  339032. The authors
acknowledge Alexandros Alexakis for very useful discussions and
comments.

\section*{Authors contribution statement}
All the authors were involved in the preparation of the manuscript.
All the authors have read and approved the final manuscript.

\bibliographystyle{unsrt}
\bibliography{biblio}

\end{document}